\documentclass[a4paper]{article}
\usepackage[utf8]{inputenc}
\usepackage{anysize}
\usepackage{graphicx}
\usepackage{mathtools}
\usepackage{amsmath}
\usepackage{bm}
\usepackage{enumerate}
\marginsize{2.85 cm}{2.8 cm}{2.8 cm}{2.8 cm}

\title{\textbf{On the number of proper $k$-colorings in an $n$-gon}}
\author{Shantanu Chhabra\\
Delhi Public School, RK Puram\\
New Delhi 110 022\\India\\
\texttt{shantanuchhabra07@gmail.com}}
\date{October 24, 2013}
\begin{document}

\maketitle
\begin{abstract}
We define an $n$-gon to be any convex polygon with $n$ vertices. Let $V$ represent the set of vertices of the polygon. A “proper” $k$-coloring refers to a function, $f$: $V$ $\rightarrow$ $\{1, 2, 3, … k\}$, such that for any two vertices $u$ and $v$, if $f(u)=f(v)$, $u$ is not adjacent to $v$. The purpose of this paper is to develop a recursive algorithm to compute the number of proper $k$-colorings in an $n$-gon. The proposed algorithm can easily be solved to obtain the explicit expression. This matches the explicit expression obtained from the popular conventional solutions, which confirms the correctness of the proposed algorithm. Often, for huge values of  $n$ and $k$, it becomes impractical to display the output numbers, which would consist of thousands of digits. We report the answer modulo a certain number. In such situations, the proposed algorithm is observed to run slightly faster than the conventional recursive algorithm. 
\end{abstract}

\smallskip
\noindent

\section{Introduction}
The $k$-colorability problem\cite{kiakai}, also known as the Chromatic Number problem or the Graph coloring problem, is one of Richard M. Karp’s 21 NP-complete problems \cite{karp}. The $k$-colorability problem is typically defined as follows: A graph is $k$-colorable if each vertex of the graph has a color different from those of its neighbors (adjacent vertices) given that we can use at most $k$ colors to color all vertices of the graph. The problem in consideration is a special case of the $k$-colorability problem in that the graph is a convex polygon with $n$ vertices ($n$-gon) and we aim at computing the number of proper $k$-colorings of an $n$-gon. Formally speaking, if $V$ represents the set of vertices of an $n$-gon, we aim at computing the number of possible functions, $f$: $V$ $\rightarrow$ $\{1, 2, 3, … k\}$, such that for any two vertices $u$ and $v$, if $f(u)=f(v)$, $u$ is not adjacent to $v$ \cite{louis}. In this paper, I propose a new implicit recurrence relation to compute the number of proper $k$-colorings of an $n$-gon. I propose a linear homogeneous recurrence relation of order $2$ for the number of proper $k$-colorings of an $n$-gon. This algorithm is developed by representing the number of proper $k$-colorings for an $n$-gon as a function of the number of proper $k$-colorings for an $(n-1)$-gon and an $(n-2)$-gon. Although the explicit solution can be implemented to run in $O(\log_2n )$, the proposed algorithm is a new implicit solution that is as efficient as the conventional implicit algorithm, i.e. both algorithms run with $\Theta(n)$ complexity.  However, for extremely large values of $n$ and $k$, the output has to be reported modulo a smaller number. With this constraint, both the proposed algorithm and the old algorithm were run on two different systems: a 64-bit Windows 7 Intel CORE i7 system and a 64-bit Ubuntu 13.10 Intel CORE i3 Sandybridge system. It was found that the proposed algorithm executed faster than the conventional algorithm on both systems in all cases.

\section{Preliminaries}
\textbf{Asymptotic Notations\cite{tim}:} Asymptotic notations can be referred to as the vocabulary for analysis of algorithms. The gist behind using asymptotic analysis for the running time of an algorithm is that asymptotic analysis helps ignore low-level details of architecture, compiler and focuses on the more important high-level reasoning while determining the efficiency of algorithms.
\\
For the purpose of this paper, I will elaborate on three fundamental asymptotic notations to make the content in this paper easier to understand: \\
\begin{enumerate}
\item \textbf{Big Oh Notation:} The Big Oh notation is represented as $O(f(n))$ where $f(n)$ is a function on the input size. The Big Oh notation is used to provide an upper bound to the running time of an algorithm in terms of a constant multiple of a function of the input size. Let $n$ be the input size and let $T(n)$ be a function on $n$. In the context of this paper, $T(n)$ will be the worst case running time of the algorithm. $T(n)$ is said to be $O(f(n))$, if there exist constants, $c$, $n_0 > 0$ such that $T(n) \le c.f(n)$ for all $n \ge n_0$.
\item \textbf{Big Omega Notation:} The Big Omega notation is represented as $\Omega(f(n))$ where $f(n)$ is a function on the input size. The Big Omega notation is used to provide a lower bound to the running time of an algorithm in terms of a constant multiple of a function of the input size. Let $n$ be the input size and let $T(n)$ be a function on $n$. In the context of this paper, $T(n)$ will be the worst case running time of the algorithm. $T(n)$ is said to be $\Omega(f(n))$, if there exist constants, $c$, $n_0 > 0$ such that $T(n) \ge c.f(n)$ for all $n \ge n_0$.
\item \textbf{Big Theta Notation:} The Theta notation is represented as $\Theta(f(n))$ where $f(n)$ is a function on the input size. The Theta notation is used to provide both a lower bound and an upper bound to the running time of an algorithm in terms of constant multiples of a function of the input size.Let $n$ be the input size and let $T(n)$ be a function on $n$. In the context of this paper, $T(n)$ will be the worst case running time of the algorithm. $T(n)$ is said to be $\Theta(f(n))$, if and only if $T(n) = O(f(n))$ and $T(n) = \Omega(f(n))$, i.e. $\exists$ constants, $c_1$, $c_2$ and $n_0$ such that $c_{1}f(n) \le T(n) \le c_{2}f(n)$ $\forall n \ge n_0$. 
\end{enumerate}

\noindent
\textbf{Binary Exponentiation:} Binary Exponentiation, also known as Exponentiation by Squaring, is a Divide and Conquer technique used to compute positive integer powers of a number. This means that the problem with a large input size is divided into sub-problems of smaller size. The solutions to these smaller problems are further combined to obtain the solution to the big problem. The recursive algorithm is as follows:
\[ x^n = \Bigg\{ 
  \begin{array}{l l}
    (x^2)^\frac{n}{2} & \quad \text{if $n$ is even}\\
    x(x^2)^\frac{n-1}{2} & \quad \text{if $n$ is odd}
  \end{array} .\]

The following C++ function implements the Binary Exponentiation recursive algorithm:
\begin{verbatim}
inline long long binexpo(int x, int n) 
{
    if(!n) return 1;
    else if(n%2)
        return x*binexpo(x, n-1);
    else if(n%2==0)
        return binexpo(x*x, n/2);
}
\end{verbatim}

\textit{Analysis:} The Binary Exponentiation algorithm runs with complexity $O(\log_{2}n)$. This is because it takes at most $\log_{2}n$ steps for the algorithm to output the final answer. \\
Throughout this paper, for the sake of convenience, I will use $\log n$ to denote $\log_{2}n$.
\\

\noindent
\textbf{Solving a Linear Homogeneous Recurrence relation of order $2$ \cite{zhang}:} We use the method of characteristic roots to obtain the explicit formula of a linear homogeneous recurrence relation (with constant coefficients). Let the recurrence relation be:\\
\centerline{$x_{n+2} = px_{n+1} + qx_n, n = 0, 1, 2, ...; p, q$ are constants and $q \not= 0$}
If the geometric sequence $\langle r^n \rangle$ ($r\not=0$) is a solution of the recurrence relation, we have $r^{n+2} = pr^{n+1} + qr^n$, i.e. $r$ is a root of the following quadratic equation: \\
\centerline{$r^2 = pr + q$}
This quadratic equation is called the characteristic equation of the sequence $\langle x_n \rangle$. The roots of the characteristic equation are called the characteristic roots of the sequence, $\langle x_n \rangle$. Conversely, if $r$ is a root of the characteristic equation, then the geometric sequence $\langle r^n \rangle$ is a solution of the recurrence relation. \\
If the two roots, $r_1$ and $r_2$ of the characteristic equation are distinct, then $\langle r_{1}^n \rangle$ and $\langle r_{2}^n \rangle$ are the solutions of the recurrence relation and for any constants, $C_1$ and $C_2$, $\langle C_1r_{1}^n + C_2r_{2}^n\rangle$ is also a solution of the recurrence relation. \\
If the initial values, $x_1 = a$ and $x_2 = b$ are given, the values of $C_1$ and $C_2$ are determined uniquely by the following system of equations: \\
\centerline{$C_1 + C_2 = a$} \\
\centerline{$C_1r_1 + C_2r_2 = b$} \\
Therefore, we get the unique solution, $x_n = C_1r_{1}^n + C_2r_{2}^n$ of the recurrence relation with the initial values $x_1 = a$ and $x_2 = b$. \\

\noindent
\textbf{Modulo operation} When we divide two integers, say $A$ by $B$, we obtain a quotient, $Q$ and a remainder $R$. The modulo operation reports $R$ for two input integers, $A$ and $B$.


\section{Conventional algorithm }
An $n$-gon is typically a convex polygon with $n$ vertices. As mentioned earlier, $V$ denotes the set of vertices of an $n$-gon. A function $f$ : $V$ $\rightarrow$ \{$1$, $2$, $3$, ... , $k$\} is called a proper $k$-coloring, if $f(u)$ $=$ $f(v)$ implies that $u$ is not adjacent to $v$. Our task is to compute the number of proper $k$-colorings for an $n$-gon. I discuss the following solution. \\
\\
\textbf{Using Recurrence Relation \cite{titu}} \\

\centerline{$ p(n,k) = k(k-1)^n - p(n-1, k) $ \cite{titu} }
where $p(n, k)$ denotes the number of proper $k$-colorings in an $n$-gon \\
\textit{Analysis:} This recurrence, if implemented using a program for large values of $n$ and $k$, will run with $\Theta(n)$ complexity. \\
The following C++ code implements the conventional algorithm.

\begin{verbatim}
#include <iostream>
#include <conio.h>
using namespace std;

inline long long binexpo(int x, int n) 
{
    if(!n) return 1;
    else if(n%2)
        return x*binexpo(x, n-1);
    else if(n%2==0)
        return binexpo(x*x, n/2);
}

int main()
{
    long long int n, k;
    static int p[n];
    cin >> n >> k ;
    p[3] = k*(k-1)*(k-2);
    long long int P = binexpo(k-1, 3);
    for (int i=4; i<=n; i++)
    {
        p[i] = k*P - p[i-1];
        P*=(k-1);
    }
    cout << p[n] << endl ;
    getch();
    return 0;
}
\end{verbatim}


\section{Proposed algorithm}
We define $g(n,k)$ to be the number of proper $k$-colorings of an $n$-gon. \\

\textit{Claim:} $g(n,k) = (k-2)g(n-1, k) + (k-1)g(n-2, k)$ \\

\textit{Proof:} The vertices of the $n$-gon are numbered from 1 to $n$. A function $f$ : $V$ $\rightarrow$ \{$1$, ... , $k$\} is called a proper $k$-coloring, if $f(u)$ $=$ $f(v)$ implies that $u$ is not adjacent to $v$. I start coloring the vertices from Vertex 1. I color Vertex $2$ such that $f(2)$ $\not=$ $f(1)$. I keep coloring vertices such that $f(r)$ $\not=$ $f(r-1)$ $ \forall$ $ r$ $ \le $ $n-2$. When coloring Vertex $(n-1)$, two cases arise \\

\textbf{Case-I:} We color Vertex $n-1$ such that $f(n-1) \not= f(1)$ \\
In this case, we're left with $ k-2 $ possibilities for Vertex $n$. Moreover, Vertex $1$ to $n-1$ have been colored such that adjacent pairs of vertices are differently colored and Vertex $1$ and Vertex $n-1$ are differently colored. This reduces to $g(n-1, k)$. Thus, from Case-I, by multiplying the number of ways to color Vertex $n$ with $g(n-1,k)$, I get the first term of my recurrence, that is $(k-2)g(n-1,k)$. \\

\textbf{Case-II:} We color Vertex $n-1$ such that $f(n-1) = f(1)$ \\
In this case, we're left with $ k-1 $ possibilities for Vertex $n$. Moreover, Vertex $1$ to $n-2$ have been colored such that adjacent pairs of vertices are differently colored and Vertex $1$ and Vertex $n-2$ are differently colored. This implies because if $f(1) = f(n-1)$ and $f(n-1) \not= f(n-2)$, then $f(1) \not= f(n-2)$ This reduces to $g(n-1, k)$. Thus, from Case-II, by multiplying the number of ways to color Vertex $n$ with $g(n-2,k)$, I get the second term of my recurrence, that is $(k-1)g(n-2,k)$. \\

The following C++ code implements the proposed recurrence algorithm.

\begin{verbatim}
#include <iostream>
using namespace std;
int main()
{
    long long int n, k;
    static int g[n];
    cin >> n >> k;
    g[2] = k*(k-1); g[3] = k*(k-1)*(k-2);
    for (int i=4; i<=n; i++)
        g[i] = (k-2)*g[i-1] + (k-1)*g[i-2];
    cout << g[n];
    return 0;
}
\end{verbatim}

\section{Verifying correctness of proposed algorithm}
\subsection*{Solving the old recurrence\cite{titu}}
$p(3,k) = k(k-1)(k-2)$ \\
$p(n,k) = k(k-1)^{n-1} - p(n-1,k)$ \\
$\implies p(n,k) = k(k-1)^{n-1} - k(k-1)^{n-2} + k(k-1)^{n-3} - ... + (-1)^{n}k(k-1)(k-2)$ \\
$\implies p(n,k) = k\frac{(k-1)^n + (-1)^{n-4}(k-1)^3}{1 + (k-1)} + (-1)^{n-3}k(k-1)(k-2)$ \\
$\implies p(n,k) = (k-1)^n + (-1)^n(k-1)^3 + (-1)^{n-1}k(k-1)(k-2)$ \\
$\implies p(n,k) = (k-1)^n + (-1)^n(k-1)[(k-1)^2 - k(k-2)]$ \\
$\implies p(n,k) = (k-1)^n + (-1)^n(k-1)$ \\
\subsection*{Solving the proposed recurrence}
The proposed recurrence is a Linear Homogeneous Recurrence relation. Therefore, we first find its characteristic equation. I use \cite{zhang} to solve my recurrence relation.\\
For $g(n,k) = (k-2)g(n-1, k) + (k-1)g(n-2, k)$, the characteristic equation will be $r^n = (k-2)r^{n-1} + (k-1)r^{n-2}$ \\
Solving the characteristic equation, we get \\
$(r - (k-1))(r + 1) = 0$ \\
$\implies r_1 = k-1$ and $r_2 = -1$ \\
From \cite{zhang}, $g(n,k) = C_{1}{r_{1}}^n + C_{2}{r_{2}}^n $ is a solution.
Now, we can find $C_{1}$ and $C_{2}$ using base-cases. \\
\centerline{$g(2) = k(k-1) = C_{1}{(k-1)}^2 + C_{2}{(-1)}^2$} \\
\centerline{and} \\
\centerline{$g(3) = k(k-1)(k-2) = C_{1}{(k-1)}^3 + C_{2}{(-1)}^3$} \\
Solving the above equations, we get $ C_{1} = 1$ and $C_{2} = k-1 $ \\
Therefore, $g(n,k) = (k-1)^n + (-1)^n(k-1)$ \\
Hence, we get the required explicit solution from the proposed recurrence as well, which confirms the correctness of the proposed recurrence. The proposed recurrence is the most efficient solution, being $\Theta(n)$ which is as efficient as the conventional recurrence algorithm.
\section{Establishing a constraint}
Often times, for large input size, the output is a number with thousands or maybe millions of digits. For instance, for $n=1000000$ and $k=1000$, the output should be $999^{1000000} + 999$, a number with nearly three million digits. Therefore, the answer is often reported modulo another number, say, $M$.\\
Both the old recurrence as well as the proposed recurrence run with $\Theta(n)$ complexity, however, it is seen that the proposed algorithm runs faster than the old algorithm when answers are reported modulo $M$.
\newpage
The following C++ code implements the old algorithm with the given constraint: \\
\begin{verbatim}
#include <iostream>
using namespace std;
const int M=10679;

static int P[N+1], T[N+1];

int main() 
{
    cin >> N >> K;
    P=(((K*(K-1))%M)*(K-2))%M;
    T=(((K*(K-1))%M)*(((K-1)*(K-1))%M)%M);
    P=(T-P+M)%M;
    
    for(int i=5; i<=N; ++i)
    {
        T=((K-1)*T)%M;
        P=(T-P+M)%M;
    }
    cout << P << endl;
    return 0;
}
\end{verbatim}

The following C++ code implements the proposed algorithm with the given constraint: \\
\begin{verbatim}
#include <iostream>
using namespace std;
const int M = 10679;
int main()
{
    long long int n, k;
    static int g[n+1];
    cin >> n >> k;
    g[2] = k*(k-1); g[3] = k*(k-1)*(k-2);
    for (int i=4; i<=n; i++)
        g[i] = ((k-2)*g[i-1] + (k-1)*g[i-2])%M;
    cout << g[n];
    return 0;
}
\end{verbatim}
\newpage
The following tables illustrate the execution times of both the old algorithm and the proposed algorithm. It can be seen that the proposed implicit recurrence algorithm runs faster than the old implicit recurrence algorithm.\\

For all of the following test-cases, $M=10679$ \\

For the Windows 7 system:

\noindent
\begin{tabular}{| c | c | c | c | c |}
\hline
\textbf{$n$} & \textbf{$k$} & \textbf{$g(n,k)$ mod $M$} & \textbf{Old Recurrence} & \textbf{Proposed algorithm}\\ \hline
 12 & 4 & 8173 & 0.001 s & 0.001 s\\ \hline
 10000 & 100 & 8014 & 0.002 s & 0.001 s \\ \hline
 100000 & 100 & 9462 & 0.009 s & 0.007 s \\ \hline
 1000000 & 100 & 3851 & 0.062 s & 0.054 s \\ \hline
 1000000 & 1000 & 7761 & 0.063 s & 0.059 s \\ \hline
 10000000 & 10 & 9279 & 0.722 s & 0.593 s \\ \hline
 10000000 & 100 & 5842 & 0.603 s & 0.598 s \\ \hline
 10000000 & 1000 & 4684 & 0.725 s & 0.610 s \\ \hline
 10000000 & 10000 & 10061 & 0.720 s & 0.588 s \\ \hline
 100000000 & 10 & 3849 & 6.163 s & 4.848 s \\ \hline
 100000000 & 100 & 10005 & 5.565 s & 4.913 s \\ \hline
 100000000 & 1000 & 3598 & 6.890 s & 4.967 s \\ \hline
 100000000 & 10000 & 6803 & 6.002 s & 5.096 s \\ \hline
 1000000000 & 10 & 1134 & 47.742 s & 30.552 s \\ \hline
\end{tabular}
\\

For the Ubuntu 13.10 system:

\noindent
\begin{tabular}{| c | c | c | c | c |}
\hline
\textbf{$n$} & \textbf{$k$} & \textbf{$g(n,k)$ mod $M$} & \textbf{Old Recurrence} & \textbf{Proposed algorithm}\\ \hline
 12 & 4 & 8173 & 0.003 s & 0.003 s\\ \hline
 10000 & 100 & 8014 & 0.003 s & 0.003 s \\ \hline
 100000 & 100 & 9462 & 0.005 s & 0.005 s \\ \hline
 1000000 & 100 & 3851 & 0.022 s & 0.016 s \\ \hline
 1000000 & 1000 & 7761 & 0.027 s & 0.017 s \\ \hline
 10000000 & 10 & 9279 & 0.186 s & 0.125 s \\ \hline
 10000000 & 100 & 5842 & 0.176 s & 0.135 s \\ \hline
 10000000 & 1000 & 4684 & 0.170 s & 0.144 s \\ \hline
 10000000 & 10000 & 10061 & 0.174 s & 0.145 s \\ \hline
 100000000 & 10 & 3849 & 1.716 s & 1.323 s \\ \hline
 100000000 & 100 & 10005 & 1.682 s & 1.434 s \\ \hline
 100000000 & 1000 & 3598 & 1.683 s & 1.403 s \\ \hline
 100000000 & 10000 & 6803 & 1.678 s & 1.403 s \\ \hline
 1000000000 & 10 & 1134 & 16.771 s & 14.778 s \\ \hline
\end{tabular}
\\

In all of the above test-cases, the proposed implicit algorithm is found to run faster than the old implicit algorithm.

\section{Conclusion}
The number of proper $k$-colorings in an $n$-gon can be efficiently computed by a linear recurrence relation. This linear recurrence relation can be solved to obtain the explicit formula, which confirms the correctness of the proposed linear recurrence relation. Furthermore, the proposed linear homogeneous recurrence, when implemented to compute the answer modulo a given number, proves to be slightly faster than the conventional recurrence algorithm.



\medskip

\begin{thebibliography}{9}
\bibitem{kiakai} Kia Kai Li. \textit{Exploring k-colorability}, 2009.
\bibitem{karp} Richard M. Karp. \textit{Reducibility among Combinatorial Problems}, University of California at Berkeley
\bibitem{louis} Math 681 Mid-term Exam, University of Louisville.
\bibitem{tim} CS161. \textit{Design and Analysis of Algorithms}, Stanford University
\bibitem{zhang} Yao Zhang. \textit{Combinatorial Problems in Mathematical Competitions}, The Methods of Finding Solutions of Recurrence Relation: The Method of Characterisitic Roots, Pages 41-42.
\bibitem{titu} Titu Andreescu, Zuming Feng. \textit{102 Combinatorial Problems from the training of the USA IMO Team}, Pages 46-47.
\end{thebibliography}
\end{document}